\documentclass[aps,twocolumn,superscriptaddress,showpacs]{revtex4-2}

\usepackage{graphicx,color}
\usepackage{dcolumn}
\usepackage{bm}
\usepackage{amsmath}
\usepackage{longtable}
\usepackage{mathtools}

\begin{document}

\preprint{}

\title{Phonon renormalization effects accompanying the 6~K anomaly in the Quantum Spin Liquid Candidate $\kappa$-(BEDT-TTF)$_{2}$Cu$_{2}$(CN)$_{3}$}

\author{Masato~Matsuura$^{*}$}
  \affiliation{Neutron Science and Technology Center, Comprehensive Research Organization for Science and Society (CROSS), Tokai, Ibaraki 319-1106, Japan}
  \email{m_matsuura@cross.or.jp}
\author{Takahiko~Sasaki}
  \affiliation{Institute for Materials Research, Tohoku University, Sendai 980-8577, Japan}
\author{Makoto~Naka}
  \affiliation{School of Science and Engineering, Tokyo Denki University, Saitama 350-0394, Japan}
\author{Jens M\"uller}
  \affiliation{Institute of Physics, Goethe-University Frankfurt, 60438 Frankfurt (M), Germany}
\author{Oliver Stockert}
  \affiliation{Max-Planck-Institut f\"ur Chemische Physik fester Stoffe, D-01187 Dresden, Germany}
\author{Andrea Piovano}
  \affiliation{Institut Laue-Langevin, 6 rue Jules Horowitz, 38042 Grenoble Cedex 9, France}
\author{Naoki Yoneyama}
  \affiliation{Graduate Faculty of Interdisciplinary Research, University of Yamanashi, Kofu, 400-8511, Japan}
\author{Michael Lang}
  \affiliation{Institute of Physics, Goethe-University Frankfurt, 60438 Frankfurt (M), Germany}
\date{\today}

\begin{abstract}
The low-temperature state of the quantum spin liquid
candidate $\kappa$-(BEDT-TTF)$_{2}$Cu$_{2}$(CN)$_{3}$
emerges via an anomaly at $T^{*}\sim6$~K. Although signatures of this anomaly 
have been revealed in various quantities, its origin has remained unclear.
Here we report inelastic neutron scattering measurements on single crystals of
$\kappa$-(BEDT-TTF)$_{2}$Cu$_{2}$(CN)$_{3}$, 
aiming at studying phonon renormalization effects at $T^{*}$.
A drastic change was observed in the phonon damping across $T^{*}$
for a breathing mode of BEDT-TTF dimers at $E=4.7$~meV.
The abrupt change in the phonon damping is attributed to a phase transition
into a valence bond solid state based on an effective model 
describing the spin-charge coupling in this dimer-Mott system.
\end{abstract}


\maketitle

Quantum spin liquids (QSLs) have been at the center of scientific attention
in the field of magnetism as a novel quantum state.
The organic charge-transfer salt $\kappa$-(BEDT-TTF)$_{2}$Cu$_{2}$(CN)$_{3}$ ($\kappa$-CN),
where BEDT-TTF is bis-(ethylenedithio)tetrathiafulvalene
C$_{6}$S$_{8}$[(CH$_{2}$)$_{2}$]$_{2}$ (ET),
has attracted attention in this area as a promising QSL candidate: 
the system is a weak dimer-Mott (DM) insulator forming a nearly isotropic two-dimensional triangular spin lattice
which lacks long-range magnetic order down to low temperatures
despite its large exchange coupling of $J\sim250$~K~\cite{Shimizu03}.
Despite the absence of a magnetic phase transition, 
anomalous behavior at low temperatures around $T^{*}\sim6$~K has been observed
in various quantities probing either spin-~\cite{Shimizu03,Pratt11,Shimizu06,Miksch21,Saito18,Isono16},
charge-~\cite{Abdel10}, lattice-~\cite{kobayashi20, Manna10, poirier14},
or composite properties thereof~\cite{SYamashita08, MYamashita09},
see also Ref.~\cite{Pustogov22}.

Recently, the discussion has taken a new twist by results of an electron spin resonance (ESR) study, 
reporting the formation of a spin-singlet state below about 6~K, 
consistent with a valence bond solid (VBS) ground state~\cite{Miksch21}, 
but contradicting with the QSL scenarios discussed so far.
Arguments in favor of a glassy VBS state have been proposed 
by Riedl \textit{et al.}~\cite{riedl19} based on their analysis of magnetic data~\cite{Isono16}. 
In addition to the signatures in the magnetic response, 
which are of moderate strength and prone to extrinsic factors~\cite{Miksch21}, 
clear evidence for the strong involvement of the lattice in the 6~K anomaly was revealed 
by measurements of thermal expansion~\cite{Manna10},
$^{63}$Cu-NQR,~\cite{kobayashi20}, and ultrasound velocity~\cite{poirier14}.
Since both the proposed QSL and VBS scenarios emerge through the 6~K anomaly, 
a clarification of its origin is key for understanding the ground state in this material.

Besides the spin- and lattice degrees of freedom, 
indications for the presence of charge degrees of freedom within the dimers 
in $\kappa$-CN and various other related DM molecular conductors were pointed out.
In $\kappa$-CN, a relaxor-type anomaly was observed in the dielectric constant below about 40~K, 
suggesting a charge disproportionation within the dimers 
and a freezing of these fluctuations on cooling towards 6~K~\cite{Abdel10, Naka10, hotta10}. 
As for the origin of these observations, conflicting results were reported 
based on measurements of charge-sensitive molecular vibrational modes.
Whereas infrared optical spectroscopy
failed to detect any clear line splitting, suggesting the absence of charge disproportionation~\cite{Sedlmeier12},
Raman scattering reveals a noticeable line broadening~\cite{yakushi15}.
A similar relaxor-type dielectric behavior was observed for the DM insulator
$\beta$'-(ET)$_2$ICl$_2$~\cite{Iguchi13}, in which an intra-dimer charge disproportionation was reported. 
In addition, clear evidence for a first-order ferroelectric transition, 
accompanied by intra-dimer charge order~\cite{Drichko2014}, 
was revealed for $\kappa$-(ET)$_{2}$Hg(SCN)$_{2}$Cl~\cite{Gati2018}. 
Moreover, for the closely related DM system $\kappa$-(ET)$_{2}$Cu[N(CN)$_{2}$]Cl ($\kappa$-Cl), 
indications for long-range ferroelectric order coinciding with antiferromagnetic ordering 
below $T_{N}=27$~K were observed~\cite{Lunkenheimer12, Lang14}.
In a recent inelastic neutron scattering (INS) study on this $\kappa$-Cl system 
it was found that the dynamics of a low-lying breathing/shearing mode of the ET dimers 
reacts sensitively to the charge- and spin degrees of freedom 
once the $\pi$-electrons become localized on the dimer site~\cite{Matsuura19}. 
Motivated by these findings, 
we report here an INS study of low-energy intra-dimer vibrational modes on $\kappa$-CN 
for probing lattice effects and their coupling to charge- and spin fluctuations 
associated with the 6~K anomaly.

\begin{figure}
\includegraphics[keepaspectratio=true,width=8.5cm,clip]{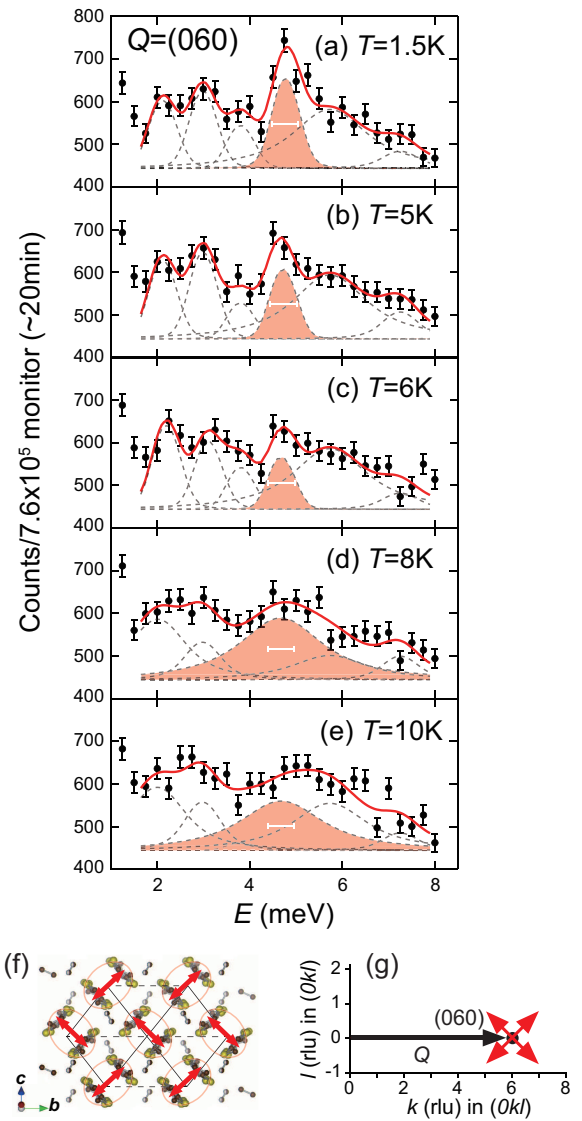}
\caption{\label{fig1} (color online)
(a)-(e) Temperature dependence of constant-Q scans at (060).
The red curves are the sum of the fits to damped
harmonic oscillator functions at $E\sim~2.0$, 2.9, 3.7, 4.7, 5.7, and 7.2~meV
convolved with the experimental resolution.
The dashed curves show the peak profiles for each mode.
The breathing mode of the ET dimer at 4.7~meV is marked in red.
The horizontal bars represent the instrumental energy resolution at $E=4.7$~meV.
(f) Top view of the ET layer.  
(g) Wave vector $Q$=(060) used for phonon measurements in the ($0kl$) scattering plane. 
Red arrows in (f) and (g) show the polarization vector $\xi$ of the breathing mode of the ET dimers,
which can be detected at (060).
}
\end{figure}

Deuterated single crystals of $\kappa$-CN were grown by utilizing an
electrochemical method~\cite{Geiser91}.
The as-grown single crystals are thin plates (thickness of $\sim0.1$~mm) with a flat
and large crystal surface (a few mm$^{2}$) parallel to the $bc$ plane 
[see Figs. S1(a), (b) in the supplemental material (SM)].
Overall, forty-seven crystals (total mass of $\sim26$ mg) were co-aligned 
within 10~degree (Fig.~S2 in SM)
according to their shape, i.e., the crystal edges including characteristic angles 
[Fig.~S1 (b) in SM]. 
The $b$- and $c$-axes were also confirmed using polarized micro infrared
reflectance spectroscopy measurements for all the single crystals. 
In neutron scattering, the phonon scattering intensity is proportional to 
($\mbox{\boldmath $Q$}\cdot \mbox{\boldmath $\xi$}$)$^{2}$
where $\mbox{\boldmath $Q$}$ is the momentum transfer between the initial and final states of the neutron,
and $\mbox{\boldmath $\xi$}$ is the polarization vector of the phonon mode.
In order to detect the breathing mode of the ET dimers
($\mbox{\boldmath $\xi$}\parallel$ [011] or [01$\bar{1}$]),
we measured phonon signals at $\mbox{\boldmath $Q$}$ = (060)
[Figs.~\ref{fig1} (f) and (g)].
INS experiments were performed using the
triple-axis spectrometer IN8 at the Institut Laue Langevin~\cite{IN8}.
The momentum transfers in this report are represented
in units of reciprocal lattice vectors $b^{*}=0.733$~\AA$^{-1}$
in the monoclinic notation.
The initial and final neutron energies were selected using
a doubly focused Cu (200) monochromator and analyzer,
which resulted in an energy resolution of 0.62~meV
at $\mbox{\boldmath $Q$} = (060)$ and $E=0$.
A pyrolytic graphite filter was placed in front of the analyzer
to suppress higher-order neutrons.

Figure~\ref{fig1} shows the temperature dependence of
the phonon spectra measured at $Q = $(060) and $T=1.5$ - 10~K.
At $T=1.5$~K, phonon peaks were observed at
$E = 2.0$, 2.9, 3.7, 4.7, 5.7, and 7.2~meV,
consistent with results of optical conductivity measurements on $\kappa$-CN~\cite{Dressel16}.
According to density-functional-theory calculations~\cite{Dressel16}, 
the peak at 4.7~meV can be assigned to an intra-dimer breathing mode.
As the temperature rises, the intensity of this mode reduces,
and all peaks become broad above 6~K.
Since the linewidth of the peak is inversely proportional to the phonon lifetime,
the broadening of these peaks indicates that
the phonon lifetime becomes substantially reduced above 6~K.

\begin{figure}
\includegraphics[keepaspectratio=true,width=8.5cm,clip]{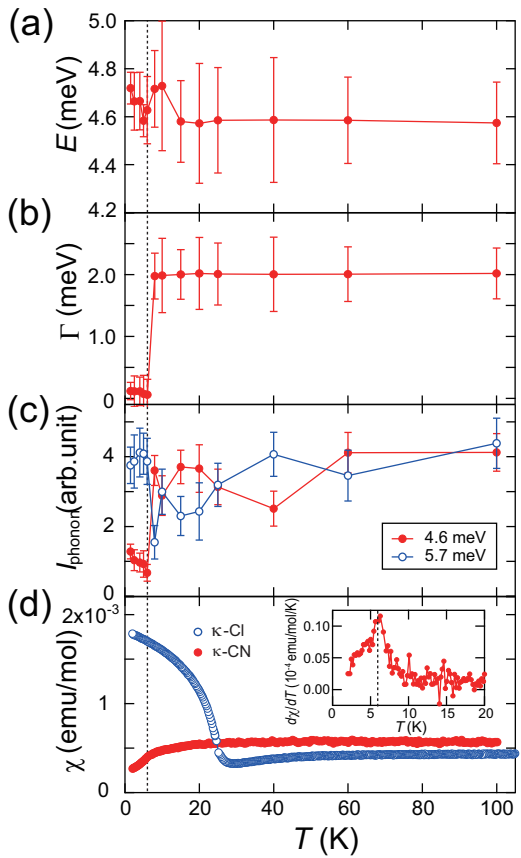}
\caption{\label{fig2} (color online)
Temperature dependence of (a) the phonon energy $E$ and
(b) the damping factor $\Gamma$ of the breathing mode of ET dimers at $E \sim 4.7$~meV.
(c) Thermal variation of the energy integrated intensity of the phonon modes ($I_\mathrm{phonon}$).
(d) Temperature dependence of the uniform magnetic susceptibility $\chi$ 
measured on a single crystal in a magnetic field of 5~T
perpendicular to the $bc$-plane and for $\kappa$-Cl taken from Ref.\cite{Matsuura19}.
Inset in (d) shows the temperature derivative of $\chi$ around the 6~K anomaly.
}
\end{figure}

The broadened phonon peaks were fitted
using the damped harmonic-oscillator (DHO) function~\cite{Gesi72}:
\begin{equation}
\mathrm{DHO}_{i}(\mbox{\boldmath $Q$},\omega)=\frac{\Gamma_{i}\hbar\omega}{[\hbar^{2} (\omega^{2}-\omega^{2}_{i})]^{2}+(\Gamma_{i}\hbar\omega)^{2}},
\label{eq1}
\end{equation}
where $\Gamma_{i}$ and $\hbar\omega_{i}$ denote
the damping factor and phonon energy of the $i$-th mode, respectively.
The phonon spectra were fitted to a constant background BG and the sum of the DHO functions
convolved with the resolution function
$R(\mbox{\boldmath $Q$},\omega)$ using
the RESTRAX simulation package~\cite{RESTRAX}:
$\sum_{i}\mathrm{DHO}_{i}(\mbox{\boldmath $Q$},\omega)\otimes R(\mbox{\boldmath $Q$},\omega)+BG$,
where $\otimes$ denotes the convolution operator.

Figures~\ref{fig2} (a)-(c) show the temperature dependence of
the DHO fitting parameters for the intra-dimer breathing mode.
The peak energy remains constant within the error bars
except for a slight hardening below $T^{*}\sim6$~K [Fig.~\ref{fig2}(a)].
Note that a similar small change ($\Delta E\sim 0.2$~meV) in the phonon energy
was observed also for the organic superconductor $\kappa$-(ET)$_{2}$Cu(NCS)$_{2}$~\cite{Pintschovius97}
upon cooling through the superconducting transition temperature,
indicating a significant electron-phonon interaction in these soft organic compounds.
The most prominent phonon anomaly in $\kappa$-CN is the change in the peak width at $T^{*}$
as observed in the phonon spectra (Fig.~\ref{fig1}) and
the corresponding reduction in the damping factor $\Gamma$ [Fig.~\ref{fig2}(b)]. 
Whereas the large damping factor above $T^{*}$ reflects an anharmonic lattice 
due to scattering processes, 
the small $\Gamma$ below $T^{*}$ suggests a freezing of these processes 
as the width becomes comparable to the instrumental resolution.
In addition to the narrowing of the phonon linewidth for $T\leq T^{*}$,
the integrated intensity ($I_\mathrm{phonon}$) of the breathing mode at $E=4.7$~meV
decreases significantly, whereas $I_\mathrm{phonon}$ is
enhanced for the $E=5.7$~meV mode [Fig.~\ref{fig2}(c)].
The intensity of the mode at $E=3.7$~meV also grows below $T^{*}$ (Fig.~\ref{fig1}).
These changes in the phonon intensities at $T^{*}$
indicate the transfer of spectral weight from the breathing modes
to other vibration modes of the ET dimers.

For a deeper understanding of the scattering processes 
involved in the phonon renormalization effects revealed here for $\kappa$-CN, 
it is instructive to compare the results with
those of the previously reported phonon study on the DM insulator $\kappa$-Cl.
Despite some structural differences to $\kappa$-CN (Fig.~S1 in SM), 
in $\kappa$-Cl, a similar increase in the damping of a phonon peak at $E = 2.6$~meV 
was observed below $T_{ins} \sim$ 50 - 60~K 
where the rapid increase in the resistivity reflects the opening of the charge gap~\cite{Matsuura19}. 
These observations suggest a close coupling of low-energy intra-dimer breathing/shearing modes 
to the $\pi$-electrons as a characteristic feature of these DM organic compounds. 

In fact, following the phonon modes to higher temperatures points to a correlation 
between the onset of phonon damping and the localization of the $\pi$-electrons on the dimer sites, 
cf.~Figs.~S4 and S5 in SM. 
Upon warming to $T = 100$~K, where $\kappa$-CN is still in its insulating, {\it i.e.}, 
high-resistance DM state, the phonon modes remain broad. 
In contrast, for $\kappa$-Cl at this temperature, 
where the $\pi$-electrons have regained a considerable degree of itineracy, 
as revealed in the resistivity, 
reflecting the closure of the charge gap, 
the spectrum exhibits well-defined phonon peaks (Figs.~S4 and S5 in SM).

These observations on $\kappa$-CN and $\kappa$-Cl suggest that 
it is the $\pi$-electrons and their localization on the dimer site 
which cause the observed phonon damping. 
According to the pseudospin-coupling model~\cite{Yamada74}, 
we expect the characteristic energy of the electrons' fluctuations 
in their charge- or spin components to be of the same size as the energies 
of the low-lying optical phonon modes, 
{\it i.e.}, $2-4$~meV. 
Once the fluctuations become frozen due to ordering in the charge- and/or spin channel, 
the phonon lifetime recovers. 
In fact, such a reduction in $\Gamma$ was observed for 
$\kappa$-Cl upon cooling below $T_\mathrm{N}=27$~K~\cite{Matsuura19} 
where magnetic order coinciding with ferroelectric order was observed~\cite{Lunkenheimer12}.

\begin{figure}
\includegraphics[keepaspectratio=true,width=8.5cm,clip]{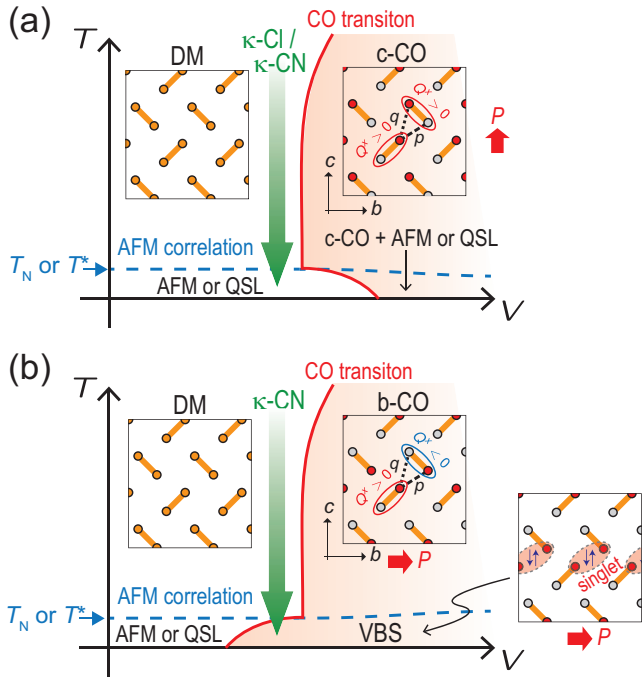}
\caption{\label{fig3} (color online)
Schematic phase diagrams of temperature ($T$) vs.
inter-dimer Coulomb interaction ($V$)
for (a) the QSL and (b) VBS states.
The broken lines represent $T_{\rm N}$ for the case of $\kappa$-Cl
whereas they correspond to a crossover for $\kappa$-CN.
Insets: spheres connected by thick orange lines represent ET dimers. 
White and red spheres correspond to 
charge-poor and charge-rich sites, respectively.
 }
\end{figure}

Similar to $\kappa$-Cl, 
the bulk of experimental findings on $\kappa$-CN indicate fluctuations and ordering phenomena 
in both the charge-~\cite{Abdel10} and spin~\cite{Miksch21} sectors. 
A minimal model that describes the coupling between the spin- and charge degrees
of freedom within the dimers has been proposed
based on the following K\"ugel-Khomskii-type
Hamiltonian~\cite{naka2016, Naka10, hotta10}:
\begin{align}
\mathcal{H}&=\sum_{\langle ij\rangle} J_{ij}\mbox{\boldmath $S$}_{i}\cdot\mbox{\boldmath $S$}_{j}
-\sum_{\langle ij\rangle} V_{ij} Q_{i}^{x} Q_{j}^{x}\\ \notag
&+2t_\mathrm{dim}\sum_{i} Q_{i}^{z}
-\sum_{\langle ij\rangle} K_{ij}\mbox{\boldmath $S$}_{i}\cdot\mbox{\boldmath $S$}_{j}
Q_{i}^{x} Q_{j}^{x}, \label{eq2}
\end{align}
where $\langle ij \rangle$ denotes the nearest-neighbor bonds;
$\mbox{\boldmath $S$}_{i}$ and $Q_{i}$ are
the spin and charge pseudo-spin operators at the $i$-th dimer unit, respectively.
The $x$ component in the pseudo-spin, $Q^x=\pm1/2$, represents the polarized states
of a hole on the dimer (cf. insets in Fig.~\ref{fig3}), and the $z$ component,
$Q^z=1/2(-1/2)$, represents a bonding (antibonding) state,
where a hole is equally distributed on the two molecules.
$J_{ij}$($>0$) is the inter-dimer exchange interaction, $V_{ij}$ is the inter-dimer Coulomb interaction,
$t_\mathrm{dim}$ ($>0$) is the intra-dimer hopping integral,
and $K_{ij}$ ($>0$) is the coupling between spins and dimer dipoles.
Owing to the spin-charge coupling $K_{ij}$, the interaction between 
the neighboring $Q^x$ is modulated by the spin-spin correlation as
$V_{\rm eff} = V_{ij} + K {\bm S}_i \cdot {\bm S}_j$.

Figure~\ref{fig3} shows schematic phase diagrams in the $V$-$T$ plane, 
deduced using the mean-field analysis of Eq.~(2)~\cite{naka2016}
where the DM ($\langle Q^x \rangle=0$) and charge order (CO)
($\langle Q^x \rangle>0$) phases compete.
In DM insulators, two patterns of CO are possible: $b$-CO and $c$-CO types,
in which polarization occurs along the $b$- and $c$-axes, respectively.
The direction of the electric polarization strongly depends on the sign of
$V_{ij}=V_p-V_q$ on the diagonal bonds, termed $p$ and $q$ in Fig.~\ref{fig3}.
In several $\kappa$-type ET compounds,
$V_p > V_q$ is obtained assuming a $1/r$-type dependence of $V_{ij}$,
which prefers the $c$-CO state~\cite{Naka10, hotta10, watanabe19}.
However, since the magnitudes of $V_p$ and $V_q$ are almost identical,
it is quite possible that other effects, e.g., due to
electron-lattice coupling, make the $V_{ij}$ effectively negative,
resulting in the $b$-CO state.
When AFM correlations develop (${\bm S}_i \cdot {\bm S}_j<0$),
$|V_{\rm eff}|$ becomes smaller for the $c$-CO ($V_{ij}>0$),
whereas $|V_{\rm eff}|$ is enhanced for the $b$-CO ($V_{ij}<0$).
Thus, due to the effects of AFM correlations, 
the $c$-CO is suppressed [Fig.~\ref{fig3}(a)],
while the $b$-CO is enhanced [Fig.~\ref{fig3}(b)],
resulting in a VBS phase, i.e., a spin-singlet state.
Therefore, depending on the type of CO, 
different $V$-$T$ phase diagrams are obtained below $T^{*}$.

In what follows, we discuss our experimental findings on phonon damping and 
its suppression below $T^{*}\sim6$~K for the two scenarios shown in Fig.~\ref{fig3}. 
In the QSL scenario ($c$-CO) [Fig.~\ref{fig3}(a)], 
these charge fluctuations become harder at low temperatures, 
resulting in a decoupling between the lattice- and charge degrees of freedom. 
For the VBS scenario ($b$-CO) [Fig.~\ref{fig3}(b)], 
spin singlets form via a spin-Peierls-like transition at $T^{*}$. 
In this case, a spin gap is opening and a hardening of charge fluctuations is expected as well.
Thus, a phonon anomaly at $T^{*}$ is expected 
through the decoupling of the lattice degrees of freedom 
from either the charge- or the charge- and spin degrees of freedom, 
corresponding to the QSL or VBS scenarios, respectively.
Then, the key difference between these scenarios
is whether the system undergoes a distinct phase transition 
at $T^{*}$ [Fig.~\ref{fig3}(b)] or a crossover [Fig.~\ref{fig3}(a)].
The present finding of an abrupt change in the phonon linewidths at $T^{*}$
is considered a strong indication for a phase transition, 
supporting the VBS scenario.

For temperatures above $T^{*}$, 
the system is situated in the DM phase near the CO phase, cf.~Fig.~\ref{fig3}. 
Upon cooling $\kappa$-CN comes closer to the CO phase boundary, 
corresponding to a softening of the intra-dimer charge fluctuations~\cite{Naka13} 
(see Fig.~S6 in SM).
This is consistent with the growth of charge fluctuations observed 
in the optical conductivity upon cooling towards $T^*$~\cite{Itoh13}. 
The distinct $c$-axis polarization of the fluctuations 
revealed in these experiments would be in support of the QSL scenario.
On the other hand, arguments in favor of a VBS scenario 
can be derived from the observation of a spin gap~\cite{Miksch21,Pustogov22} 
in combination with findings from thermal expansion measurements 
where well-pronounced signatures were observed at $T^*$~\cite{Manna10, Manna18}. 
In particular, these latter data provide strong evidence 
for a second-order phase transition at $T^{*}\sim6$~K, 
rather than a crossover, albeit with significant sample-to-sample variations 
in the shape of the anomaly~\cite{Manna18}. 
Moreover, for those crystals were the effects are most strongly pronounced, 
the phase transition anomaly shows striking similarities 
to that revealed for the spin-Peierls transition in (TMTTF)$_2$AsF$_6$~\cite{Souza09}. 
A phase transition into a VBS state, suggested by these results, 
would also be consistent with the abrupt change in the phonon linewidth, 
and the sharp feature observed in d$\chi$/d$T$ at $T^{*}$ 
[cf. inset of Fig.~\ref{fig2}(d)].

In conclusion, by studying the spectra of selected low-energy optical phonons of the quantum spin liquid
candidate $\kappa$-(ET)$_2$Cu$_{2}$(CN)$_3$ as a function of temperature,
we observe an abrupt change in the phonon linewidth at $T^{*}\sim6$~K.
We argue that the recovery of long-lived (underdamped) breathing modes of ET-dimers below 6~K
can be attributed to a cooperative phenomenon 
involving the lattice and its coupling to the charge- and 
spin degrees of freedom around 6~K. 
Our data are consistent with the formation of a VBS state below 6~K.  

We thank N. Sato and S. Sugiura
for their help in preparating the experiments.
The neutron experiments were performed with the approval of ILL (7-01-513).
This study was financially supported by
Grants-in-Aid for Scientific Research
(19H01833, 19K03723, 20H05144, and 22H04459) from the Japan Society for the Promotion of Science. 
Work at Goethe University, Frankfurt, was supported by the Deutsche Forschungsgemeinschaft
(DFG, German Research Foundation) for
funding through TRR 288 - 422213477 (project A06 and B02).

\end{document}



\title{Supplemental material\\ for\\ Phonon renormalization effects accompanying the 6~K anomaly in the Quantum Spin Liquid Candidate $\kappa$-(BEDT-TTF)$_{2}$Cu$_{2}$(CN)$_{3}$}

\author{Masato~Matsuura$^{*}$}
  \affiliation{Neutron Science and Technology Center, Comprehensive Research Organization for Science and Society (CROSS), Tokai, Ibaraki 319-1106, Japan}
  \email{m_matsuura@cross.or.jp}
\author{Takahiko~Sasaki}
  \affiliation{Institute for Materials Research, Tohoku University, Sendai 980-8577, Japan}
\author{Makoto~Naka}
  \affiliation{Waseda Research Institute for Science and Engineering, Waseda University, Shinjuku, Tokyo, 169-8555, Japan}
\author{Jens M\"uller}
  \affiliation{Institute of Physics, Goethe-University Frankfurt, 60438 Frankfurt (M), Germany}
\author{Oliver Stockert}
  \affiliation{Max-Planck-Institut f\"ur Chemische Physik fester Stoffe, D-01187 Dresden, Germany}
\author{Andrea Piovano}
  \affiliation{Institut Laue Langevin, 6 rue Jules Horowitz, 38042 Grenoble Cedex 9, France}
\author{Naoki Yoneyama}
  \affiliation{Graduate Faculty of Interdisciplinary Research, University of Yamanashi, Kofu, 400-8511, Japan}
\author{Michael Lang}
  \affiliation{Institute of Physics, Goethe-University Frankfurt, 60438 Frankfurt (M), Germany}
\date{\today}

\maketitle

\section{Supplementary Material}
\begin{figure}
\includegraphics[keepaspectratio=true,width=10cm,clip]{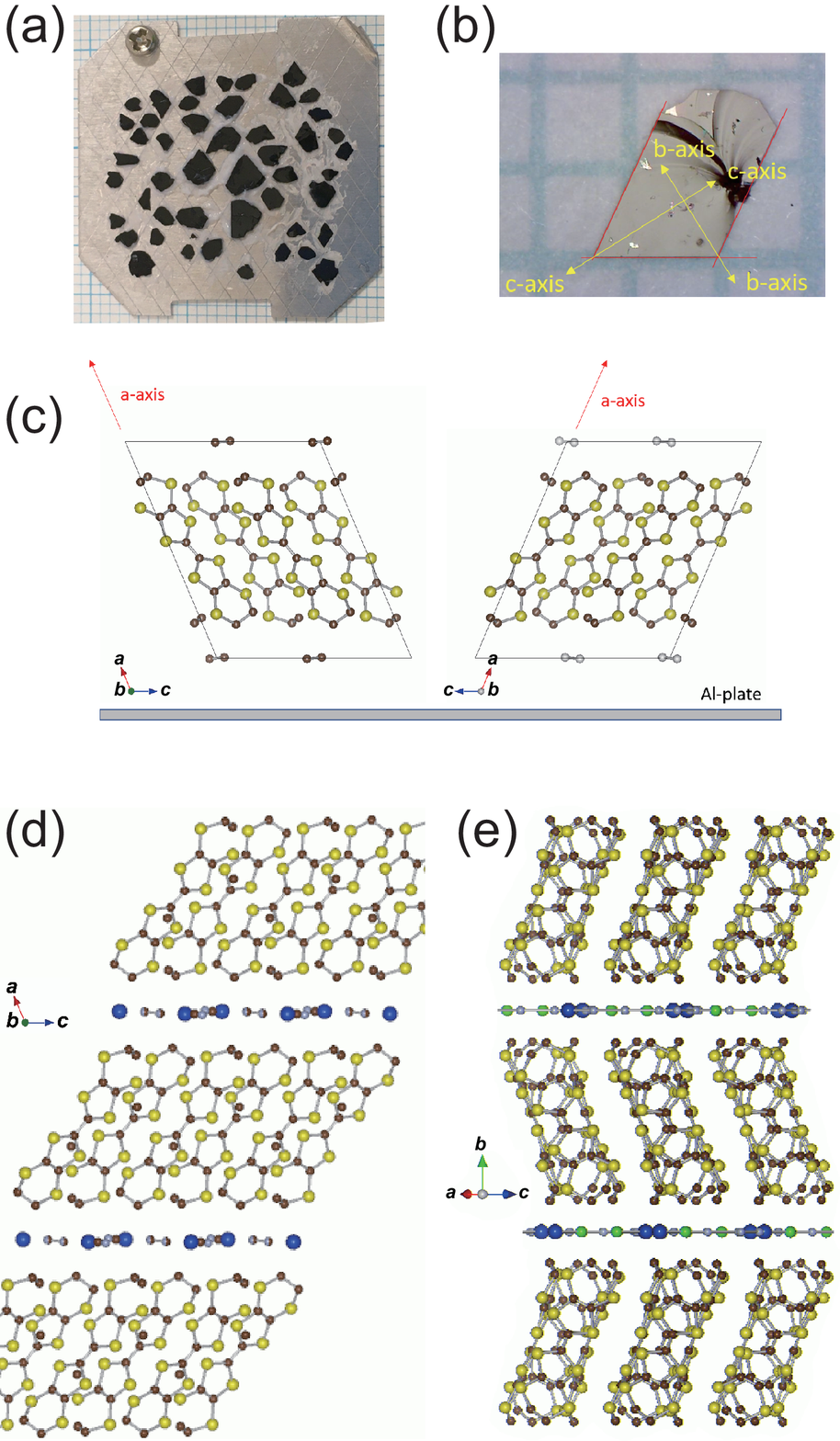}
\caption{\label{fig1} (color online)
Images of (a) co-aligned $\kappa$-(ET)$_2$Cu$_{2}$(CN)$_3$ ($\kappa$-CN) crystals 
and (b) the typical shape of the $\kappa$-CN crystal with a flat surface of the $bc$-plane. 
Crystal structure of $\kappa$-CN : (c) side view of the layered structure. 
Side view of the layered structure for (d) $\kappa$-CN and
(e) $\kappa$-(ET)$_{2}$Cu[N(CN)$_{2}$]Cl.
}
\end{figure}

Figures~\ref{fig1}(a)-(b) show photos of 
co-aligned $\kappa$-(ET)$_2$Cu$_{2}$(CN)$_3$ ($\kappa$-CN) crystals 
and the typical shape of the $\kappa$-CN crystal.
Owing to the monoclinic symmetry $P2_{1}/C$, 
with a monoclinic angle 113.42~deg and lattice parameters of 
$a=16.117$~\AA, $b=8.586$~\AA, and $c=13.397$~\AA~\cite{Geiser91},  
the $a$-axis is not perpendicular to the $bc$ plane,
and the crystals mounted on the aluminum plate 
contain two choices of the structural domains, 
according to their tilted $a$-axes [Fig.~\ref{fig1} (c)]. 
Phonon signals were measured along the $b^{*}$($=c\times a$) axis
at $\mbox{\boldmath $Q$}$=(060), 
which is common to the two tilted $a$-axes.
In each ET layer, the long ET molecular axes are inclined approximately $30^{\circ}$ 
with respect to the layer normal. 
In $\kappa$-CN with a monoclinic structure, 
ET layers of the same orientation are stacked with an anion layer in between [see Fig.~\ref{fig1}(d)]. 
In contrast, two adjacent ET layers with opposite orientations are stacked alternately 
in $\kappa$-(ET)$_{2}$Cu[N(CN)$_{2}$]Cl ($\kappa$-Cl) 
with an orthorhombic structure~\cite{Geiser91_PhysC}.
Note that the differences in the mode energies revealed 
for $\kappa$-CN and $\kappa$-Cl are likely 
due to differences in the crystal structure, resulting in a somewhat harder lattice for $\kappa$-CN.
Given that in $\kappa$-CN, the ET dimer is stacked
in the same orientation and the anion layer has a closely packed structure
[Figs.~\ref{fig1}(d) and (e)],
the phonon modes of $\kappa$-CN
are expected to be harder compared to those of $\kappa$-Cl.

\begin{figure}
\includegraphics[keepaspectratio=true,width=10.0cm,clip]{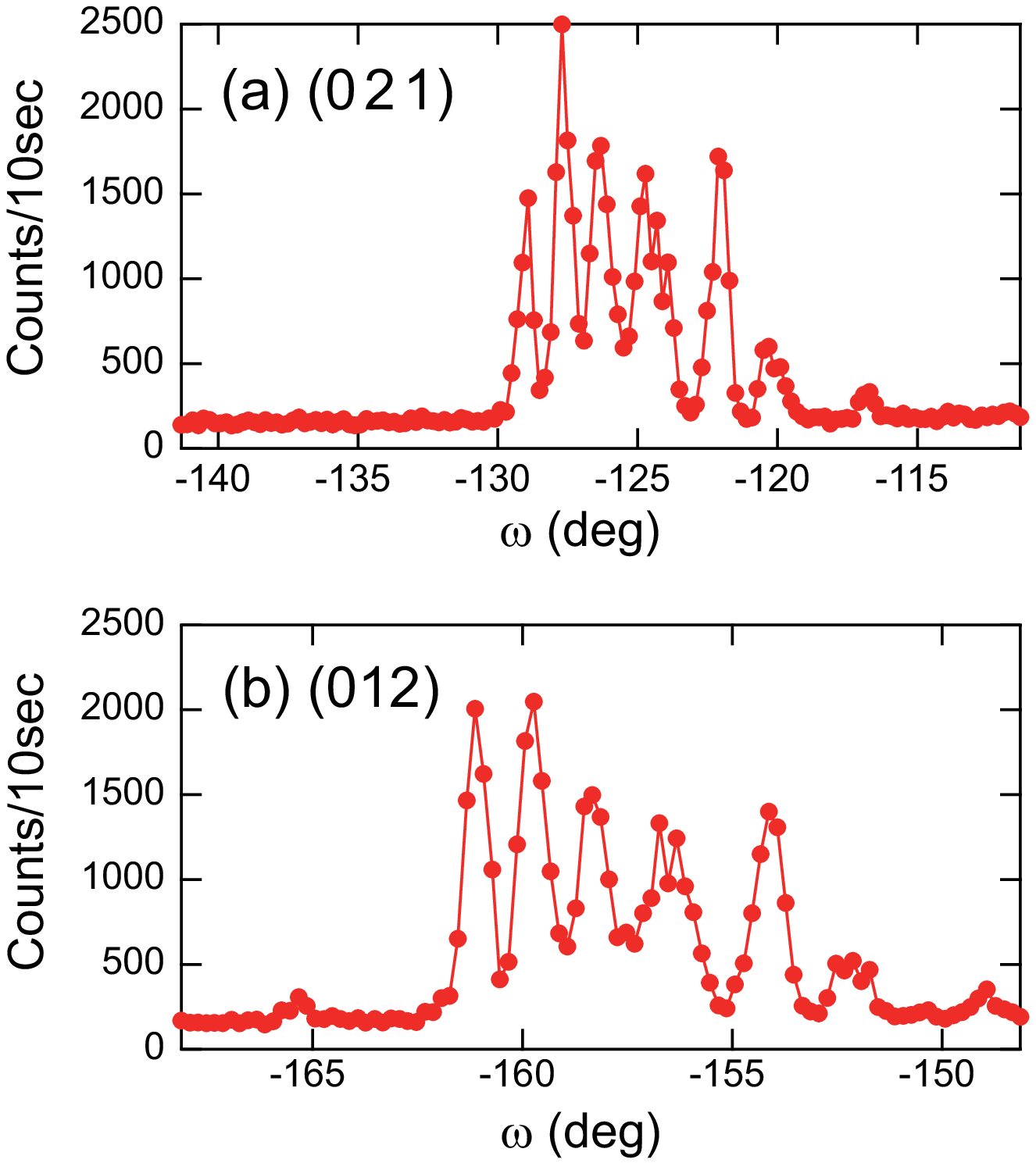}
\caption{\label{fig2}
Rocking curves of (a) (021) and (b) (012) Bragg peaks 
for the co-aligned $\kappa$-CN crystals obtained using the IN3 spectrometer.
}
\end{figure}

Figures~\ref{fig2} show rocking curves of (021) and (012) Bragg peaks
for the co-aligned $\kappa$-CN crystals. 
Most of the crystals are co-aligned within 10 degrees for both Bragg peaks.
Although the overall mosaicity of the sample 
is larger than the instrumental resolution ($\sim1$~deg), 
the main conclusion of this study is not affected
because the phonon mode at $E=4.7$~meV 
is an optical mode with no dispersion near $\Gamma$-points.

\begin{figure}
\includegraphics[keepaspectratio=true,width=13.0cm,clip]{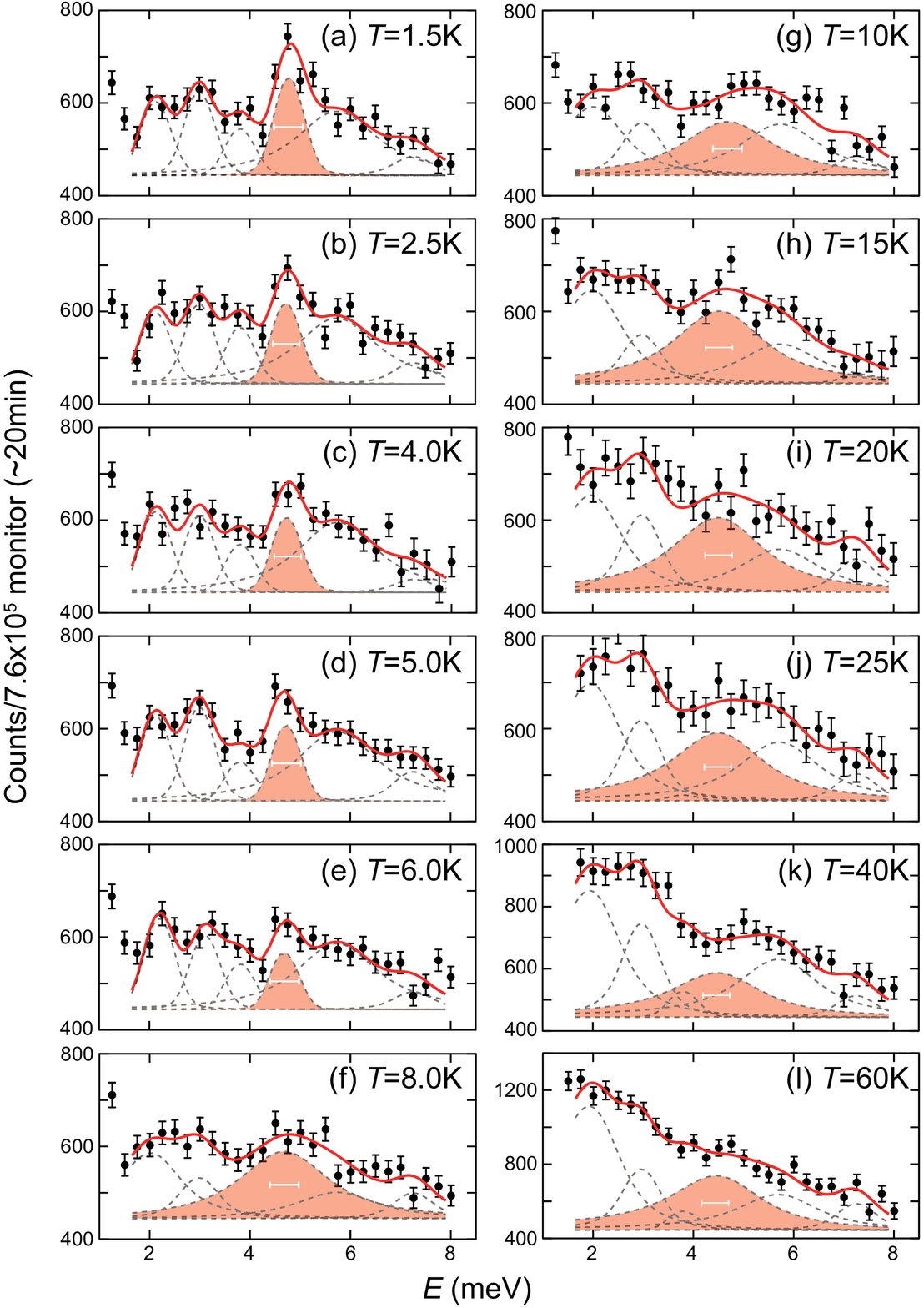}
\caption{\label{fig3}
Temperature dependence of constant-Q scans at (060).
The solid lines are fits to damped
harmonic oscillator functions at $E\sim~2.0$, 2.9, 3.7, 4.7, 5.7 and 7.1~meV
convolved with the instrumental resolution.
The dashed lines show the peak profiles for each mode.
The breathing mode of the ET dimer at 4.7~meV is represented by a red hatch.
The horizontal bars represent instrumental energy resolution at $E=4.7$~meV.
}
\end{figure}

\begin{figure*}
\includegraphics[keepaspectratio=true,width=10.0cm,clip]{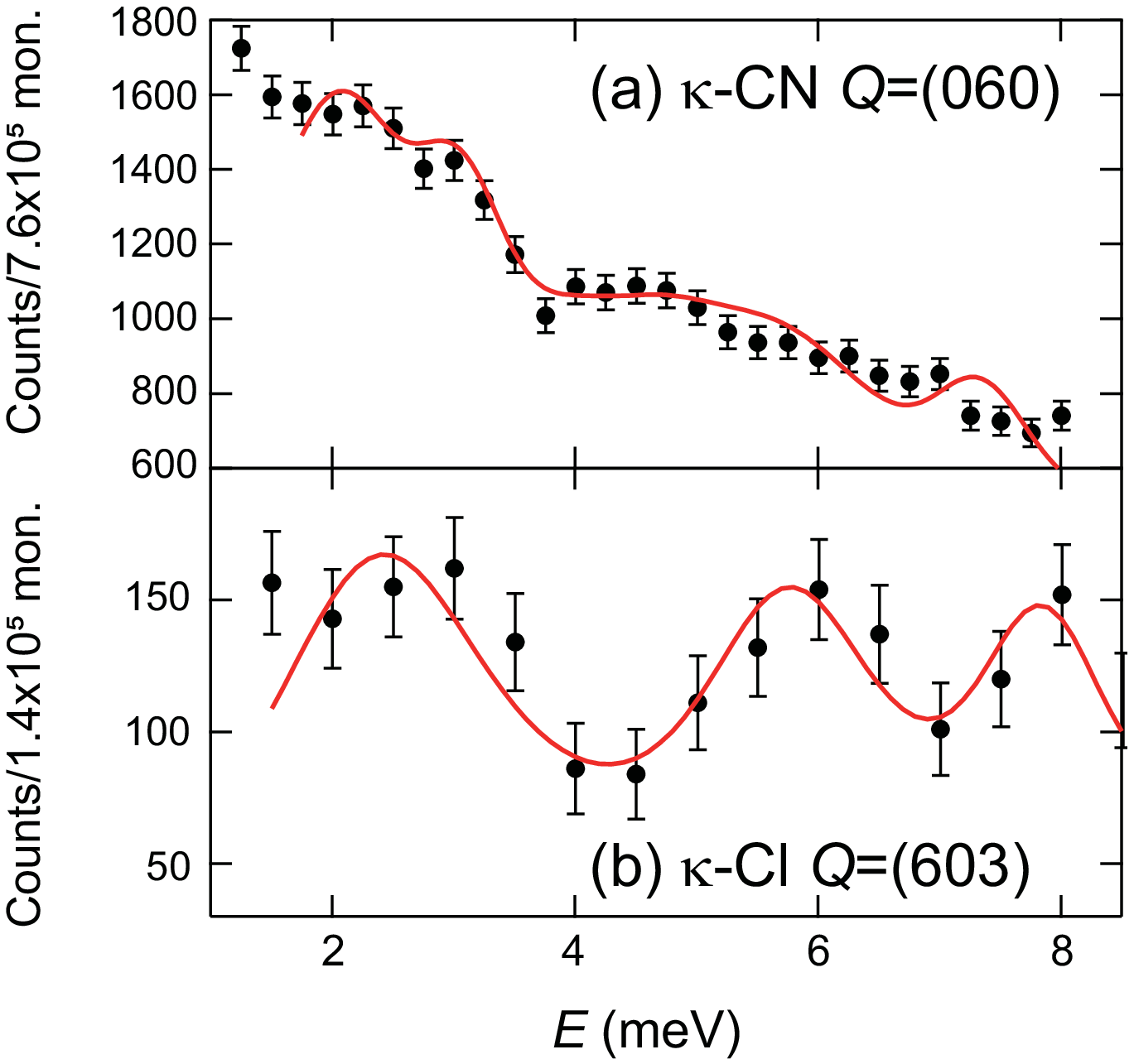}
\caption{\label{fig4} 
Phonon spectra at $T=100$~K for $\kappa$-CN and $\kappa$-Cl~\cite{Matsuura19}
which were measured at $Q=$(603) and (060), respectively.
}
\end{figure*}

Figure~\ref{fig3} shows the all temperature dependence 
for constant-Q scans at (060) for 
$\kappa$-(BEDT-TTF)$_{2}$Cu$_{2}$(CN)$_{3}$ ($\kappa$-CN),
some of which are not shown in the manuscript.
The phonon spectra for $\kappa$-CN are broad above $T=6$~K.
This is in contrast to the phonon spectra of $\kappa$-(BEDT-TTF)$_{2}$Cu[N(CN)$_{2}$]Cl
($\kappa$-Cl) showing recovery of sharp phonon peaks at $T=100$~K,
as shown in Fig.~\ref{fig4}.
The differences in phonon damping between $\kappa$-CN
and $\kappa$-Cl could be related to the itinerancy of 
$\pi$-electrons.

\begin{figure}
\includegraphics[keepaspectratio=true,width=10cm,clip]{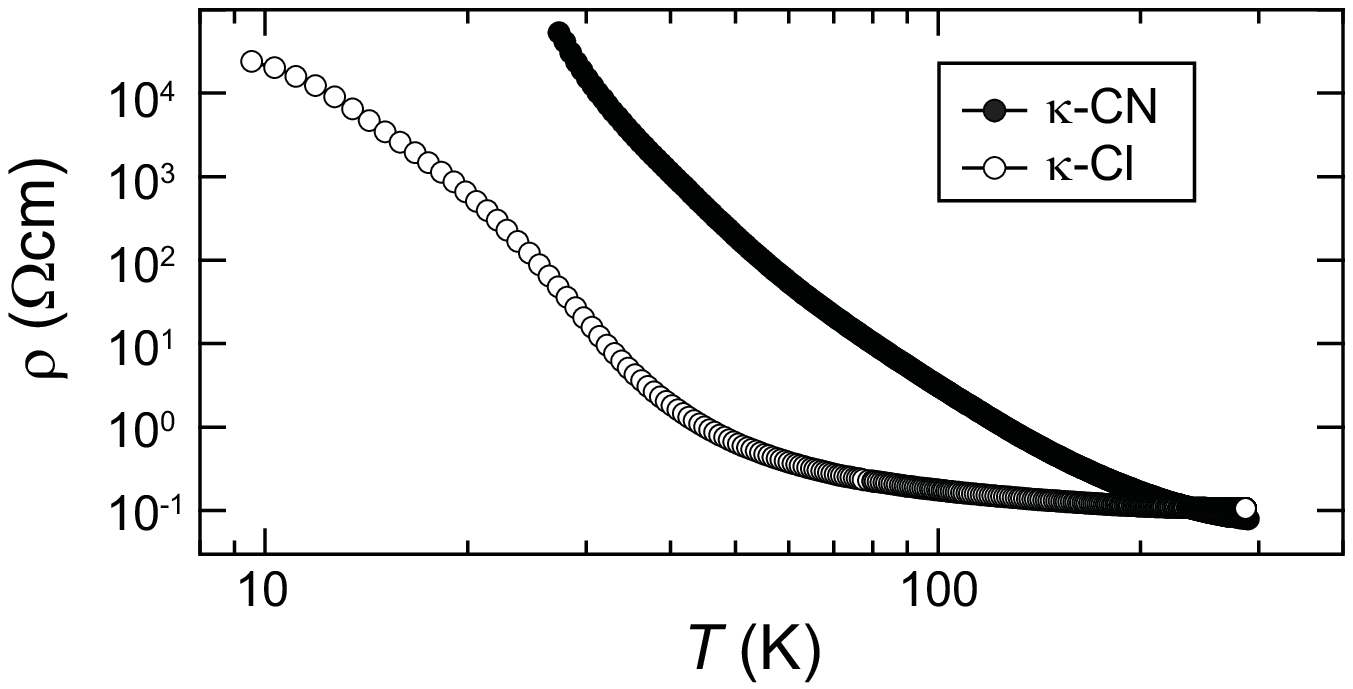}
\caption{\label{fig5} 
Temperature dependence of the electric resistivity $\rho$
for $\kappa$-CN and $\kappa$-Cl. 
A deuterized sample was used for $\kappa$-Cl
whereas a hydrogenated sample was used for $\kappa$-CN.
}
\end{figure}

Figure~\ref{fig5} shows the temperature dependence 
of the electric resistivity $\rho$ for $\kappa$-CN and $\kappa$-Cl~\cite{Matsuura19}. 
Although a hydrogenated sample was used for $\kappa$-CN,
deuterization did not affect on the resistivity in the case of $\kappa$-CN
since $\kappa$-CN is deep in the dimer Mott-insulator phase.
$\rho$ for $\kappa$-CN shows semiconducting behavior 
for all temperatures below $T=300$~K.
However, $\rho$ for $\kappa$-Cl shows
a rapid increase below $T_\mathrm{INS}\sim50-60$~K
below which the charge gap opens in optical conductivity.

%

\begin{figure}
\includegraphics[keepaspectratio=true,width=10cm,clip]{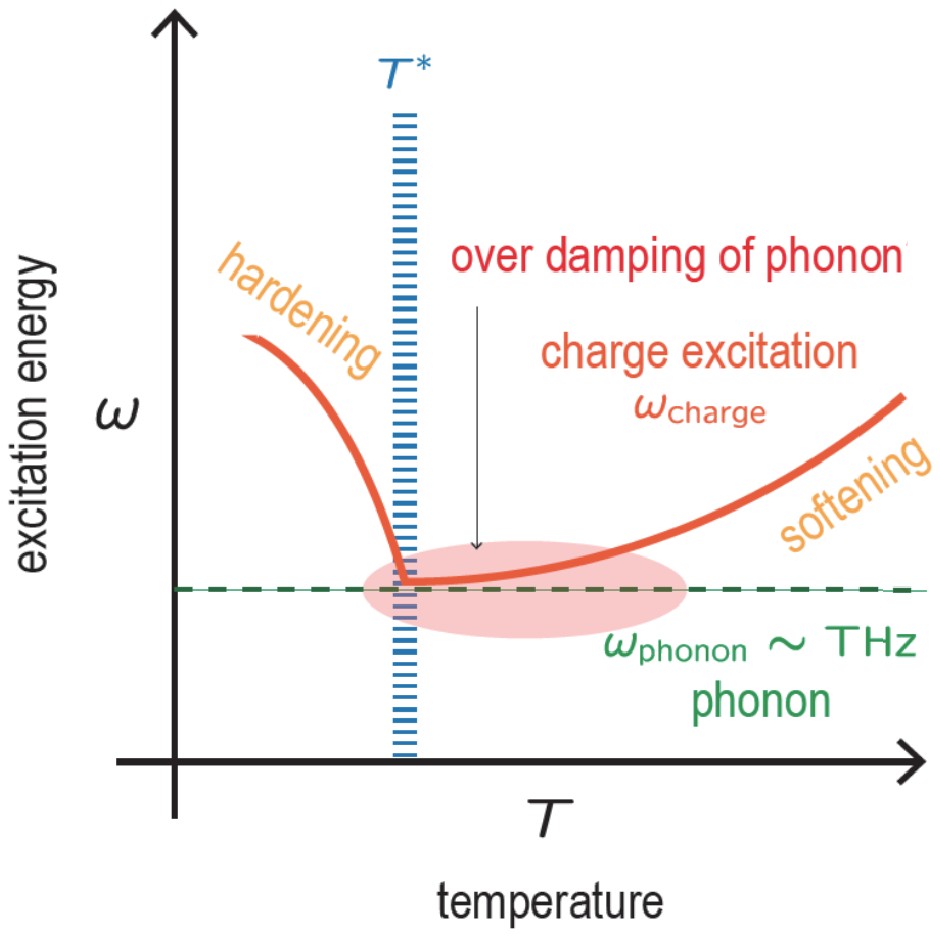}
\caption{\label{fig7}
Schematic diagram of the 
temperature dependence of the charge
fluctuations $\omega_\mathrm{charge}$ 
and coupled phonon frequency $\omega_\mathrm{charge}$.
}
\end{figure}

As the temperature decreases, 
$\kappa$-CN approaches the CO phase boundary,
and the energy of the intra-dimer charge fluctuation exhibits 
softening (Fig.~\ref{fig7})~\cite{Naka13}.
In the theoretical phase diagram for $c$-CO [Fig.~3(a) in the main text],
reentrant behavior of the CO phase boundary is observed
in the AFM/QSL phase
owing to the competition between the exclusive AFM and CO phases.
In the AFM/QSL phase, 
$\omega_\mathrm{charge}$ becomes harder as 
it shifts away from the CO phase boundary (Fig.~\ref{fig7}).
Thus, the coupling between the charge and lattice is expected to 
be weak below $T^{*}$ because $\omega_\mathrm{charge}$ and 
$\omega_\mathrm{phonon}$ are no longer close to each other.

The crystal structures in Figs.~\ref{fig1} were produced by VESTA software\cite{Momma_11}.